\documentclass{article}

\usepackage{cite}

\usepackage{graphicx}
\usepackage{amsmath, amssymb}
\usepackage{algorithmic}
\usepackage{comment}

\usepackage{xcolor}
\usepackage{algorithm}  
\usepackage{bbm}
\usepackage{enumitem}
\usepackage{authblk}
\begin{document}

\title{Capacity Expansion Planning under Uncertainty subject to Expected Energy Not Served Constraints}

\author[1]{Marilena~Zampara}
\author[2]{Daniel~\'Avila}
\author[1]{Anthony~Papavasiliou}
\affil[1]{National Technical University of Athens}
\affil[2]{Université Catholique de Louvain}

\maketitle

\begin{abstract}
We present a method for solving a large-scale stochastic capacity expansion problem which explicitly considers reliability constraints, in particular constraints on expected energy not served. Our method tackles this problem by a Lagrange relaxation of the expected energy not served constraints. We solve the relaxed formulation in an iterative manner, using a subgradient-based method. Each iteration requires the solution of a stochastic capacity expansion problem, for which we implement a subgradient decomposition scheme in a high-performance computing infrastructure. We apply the proposed methodology on the Economic Viability Assessment model that is used by ENTSO-E in the annual European Resource Adequacy Assessment, extended to include explicit reliability constraints. The approach is able to solve this model achieving a 1.3$\%$ optimality gap. We compare our approach against accounting for reliability through penalizing load shedding at VOLL, and find that the former results in 1.6$\%$ savings in total cost. We are also able to quantify the cost savings from allowing some load curtailment in the capacity planning process, which ranges from 1.6 to 6$\%$ in the cases analyzed. 

\end{abstract}


\section{Introduction}

Capacity expansion planning (CEP) involves the identification of a cost-optimal plan for investments so that the power system can meet power demand requirements subject to limitations, technical or other. A CEP, even deterministic, can be hard to solve if the level of detail is high, e.g. with extended geographic coverage, consideration of the transmission network, detailed representation of power plants and their technical limitations, and so on.


The consideration of uncertainty naturally adds further complexity to the CEP. Firstly, stochastic formulations and relevant solution methods are inherently more complex. Secondly, the incorporation of a large number of scenarios (which is necessary for adequately capturing uncertainty) and the corresponding increase in decision variables significantly increases the size of the models, rendering them computationally intractable. There is thus a trade-off between modeling accuracy and tractability \cite{Conejo_book_2016}.


The strategies for tackling the stochastic CEP can generally be classified into two major categories: a) applying some form of scenario reduction \cite{feng2013,hilbers2019,sun2019}, which tips the scale towards tractability and away from accuracy, and b) implementing a decomposition technique \cite{bloom1982long,sirikum2007new,soares2022integrated,Da_Costa,davila}, tipping the scale the other way around. The two strategies can of course be combined  \cite{roh2009market}.


The stochastic CEP problem involves a natural sequence of events that make it decomposable; first, investment decisions are made, prior to the realization of uncertainty. Then, uncertainty is revealed, e.g. wind generation is low or high, and the system responds accordingly by deciding which capacities to operate and by how much. We thus have a natural two-stage structure. The first stage involves an investment problem, and the second stage several operation problems for a given level of investment, each for every considered scenario. This two-stage structure allows for implementing decomposition methods \cite{birge2011introduction}. Currently, the literature includes algorithms based on decomposition methods that are able to solve even very large stochastic CEP problems \cite{davila}.


When reliability constraints are introduced to the stochastic CEP, decomposing the problem becomes quite challenging. A reliability constraint would be in the form of e.g. expected load shedding or hours with load shedding being less than or equal to a limit. This means that a reliability constraint bundles all scenarios together in an explicit way (as opposed to the implicit coupling that is implied by first-stage decisions), which does not allow for directly decomposing the problem by scenario. This is why an explicit consideration of reliability constraints in the stochastic CEP is considered as an “extremely hard” task \cite{Da_Costa}.


Regardless of its complexity, the task is highly relevant in practical applications; the vast penetration of renewable resources has created a need to account for uncertainty in a systematic way. At the same time, reliability considerations are an integral part of capacity expansion planning. 
This is reflected in the European Resource Adequacy Assessment (ERAA), the annual analysis carried out by the European Network of Transmission System Operators for Electricity (ENTSO-E) of whether European power systems will have sufficient resources to meet electricity demand in a ten-year time frame \cite{ERAA_website}. The first sentence in the description of the ERAA methodology \cite{ERAA23_method} highlights its stochastic scope: “Adequacy studies aim to evaluate a power system’s available resources … under a variety of scenarios.” Nevertheless, computational feasibility issues manifest, and mold the methodological approach of the study. 


In the following, we will briefly present the methodology of the ERAA and discuss challenges involved. We will then discuss how stochastic capacity expansion with reliability is addressed in the literature.

\subsection{The European Resource Adequacy Assessment}
The following will be based on the ERAA studies from 2021 to 2023, as the ERAA 2024 study has not been published at the time that this paper is written. 

\subsubsection{Methodology} \label{ERAA_method}
The ERAA is a two-step process; the first step is the “Economic Viability Assessment” (EVA), that investigates “how capacity will evolve”. The second step is the “Adequacy Assessment”, that investigates whether “capacity will be enough”.


As its name suggests, the EVA assesses the economic viability of existing capacities in an energy-only market (EOM), as well as the ability of the EOM to deliver new investments. The EVA is essentially a least-cost capacity expansion problem, yielding decisions on the investment and retirement of capacities. In the most recent versions of the study (2022 and 2023), the EVA model is formulated over a lookahead of ten years. 
Reliability standards are not explicitly considered in this step. However, shedding of load is limited by introducing a cost of curtailed energy in the objective function. In the most recent versions of the ERAA this cost has been set equal to the price cap of the single day-ahead market coupling (SDAC), while in the ERAA 2021 the VOLL was used. In both cases this cost of curtailed energy is defined exogenously.


The Adequacy Assessment part of ERAA considers the capacities projected with the EVA and adopts a Monte Carlo approach for calculating reliability indices. The comparison between these calculated reliability indices and national reliability standards identifies the areas with a future risk of capacity adequacy.


The ERAA accounts for uncertainty in climate conditions (which affects renewable energy generation and demand) and in the occurrence of unscheduled outages. Variability in climatic conditions is represented in 35 climatic scenarios. However, since the EVA is a very large  capacity expansion problem, it becomes hard  to incorporate many scenarios. For this reason, the most recent versions of ERAA adopt a scenario clustering approach for the EVA, developed by \cite{daniel_phd}, where the number of studied climatic scenarios is reduced from 35 down to 3. Moreover, unscheduled outages are accounted for by derating capacities. Uncertainty is fully accounted for in the adequacy assessment step.

\subsubsection{Challenges}\label{ERAA_chal}

Let us focus on challenges that we identify in the ERAA methodology. The first one regards the limited number of scenarios that are used in the EVA, which has been criticized as a significant shortcoming of the assessment (item (23) of \cite{acer_note}). The use of a limited number of scenarios reduces the computational effort for the EVA, however it introduces the challenge to identify a proper scenario reduction methodology.  
In the ERAA 2023, two different clustering methods are implemented; the selection of different methods and the comparison of results add complexity to the assessment. 


Another challenge that we identify in the EVA study is the determination of the level of the market price cap. The EVA text \cite{ERAA23_method} highlights the importance of setting the price cap properly: “The value of the price cap is of first order of importance when assessing energy market viability of resource capacities.” Indeed, a high price cap drives higher levels of investment, while a lower cap results in lower investments and more retirements. The price cap acts as an implicit representation of the reliability targets of every country considered in the EVA. Moreover, it has an institutional significance. In particular, ACER has issued specific directions  \cite{acer_pricecap} with regards to setting market price caps for the SDAC, based on the observed day-ahead market prices. The ERAA 2023 attempts to implement these directions, albeit an exact implementation is not possible with the modeling tools used for the EVA \cite{ERAA23_method}.      


Finally, there is an issue of consistency between the two steps of the ERAA. ENTSO-E supports that each of the two steps of the ERAA serves a different purpose (item (34) in \cite{acer_note}), nevertheless, it has acknowledged that "ideally, the EVA and risk models are performed in an integrated model". This issue has been a significant motivation for the work presented in this paper.   

\subsection{Stochastic CEP with Reliability Constraints in the Literature}

We begin the literature review by defining common reliability metrics utilized in the context of CEP. We can generally identify two types of metrics; those that evaluate frequency and those that evaluate severity. The former group includes the loss of load expectation or probability (LOLE or LOLP), which measure the incidents (hours) of load curtailments. The latter includes the expected energy (or power) not served (or supplied) (EENS or EPNS) and the loss of energy probability (LOEP), which measure the amount of load curtailment. 
Note that the use of LOLE/LOLP constraints in a CEP problem would in principle require counting load shedding incidents (i.e. possibly integer variables), which would render the problem non-convex and hence computationally complex. This could possibly rationalize why, in the literature of CEP with reliability constraints, we see the use of EENS/EPNS/LOEP metrics more often than LOLE/LOLP.


Let us now proceed with a review of the methodologies for tackling the stochastic CEP with reliability constraints. In the literature, the majority of works implement some form of enhanced Bender’s decomposition or combine Bender's decomposition with other techniques. \par
An early example is \cite{roh2009market}, that solves a market-based stochastic CEP with LOEP constraints. It proposes to split the problem into three stages, investment, reliability, and operation, and implements Bender’s decomposition where the master problem corresponds to the investment stage, and where the solution of reliability and operation problems provide cuts for the investment problem. 


Decomposition into three levels (investment, reliability, and operation) is adopted also in \cite{robust2016} to solve a least-cost capacity expansion problem that satisfies a constraint on EENS. This work combines Bender’s decomposition with robust optimization; the reliability subproblem is solved only for the worst-case realization of uncertainties, which limits the number of cuts and thus allows for solving large models. 


The same tri-level Bender’s decomposition structure is employed in \cite{Da_Costa} to solve a least-cost capacity expansion problem satisfying EENS, as well as CVaR constraints, a metric borrowed by the financial sector. Stochastic dual dynamic programming \cite{Pereira1991} is used to solve the operation subproblem and Monte Carlo simulations for the reliability subproblem. 


In other works, the consideration of reliability constraints is implicit through the incorporation of a relative cost term in the objective function. This approach is adopted in the EVA study, as already discussed in \ref{ERAA_method}. Explicit reliability constraints are omitted, which simplifies the models considerably. The challenge in this case is the proper definition of the cost of load loss. Theoretically, if the cost of load loss is properly defined, then this would be sufficient to yield the same level of investment as a specific EENS constraint would. This argument is based on strong duality; if we implement Lagrange relaxation in a problem with EENS constraints, then the violation of said constraint is penalized in the objective function by its Lagrange multiplier (or shadow price) \cite{boyd2004convex}. This is equivalent to pricing involuntary load shedding at a value of lost load. Two examples that account for reliability in this implicit manner are \cite{jirutitijaroen2008} and \cite{aghaei2014}.


We will conclude this review by addressing the question: if we were to add explicitly reliability constraints to the EVA model of ERAA, could we implement the aforementioned approaches? 


Let us first examine size considerations, as the EVA involves a pan-European, thus very large, model.  
Out of the methodologies mentioned, the robust method of \cite{robust2016} is applied to the largest model (73-bus system with 96 units and 120 lines), which is smaller than the size of the EVA (56 nodes with 1260 units and 664 lines). The exact size limitations of the presented methodologies have not been examined. However, all presented methodologies are based on Bender's decomposition, and the work of \cite{davila} demonstrates the shortcomings of Bender's decomposition in solving a problem that is as large as the EVA. Da Costa et al. \cite{Da_Costa} also comments on the fact that the performance of Bender's decomposition deteriorates as the problem increases in scale.


Let us now comment on aspects other than size limitations. Roh et al. \cite{roh2009market} also formulate a capacity expansion model subject to LOEP constraints. However, the scheme proposed by the authors relies on a heuristic iterative procedure, without convergence guarantees. 
Moreover, the computational tractability of the approach is ensured through scenario reduction. It would thus not resolve the challenge of incorporating multiple scenarios in the EVA (discussed in section \ref{ERAA_chal}). For the same reason, \cite{robust2016} would also be unsuitable, as it focuses on reliability under a single scenario. Finally, in the approach of \cite{Da_Costa}, the model analyzed has some differences with the EVA model. There is no consideration of the network and this allows the authors to easily approximate load shedding, however network constraints would render this approximation invalid.

\subsection{Contribution of the Present Work}

The work presented in this paper aims at solving a large stochastic CEP with EENS constraints. The motivation for the work are the challenges that have been encountered in the ERAA study, and specifically the separate assessment of capacity expansion (EVA) and adequacy. We will thus specifically work on a model similar to the EVA, to which we add explicit reliability constraints. The reliability metric that we utilize is EENS. We choose EENS over LOLE/LOLP because, as already discussed, the latter involves counting incidents, which would render our problem non-convex and effectively computationally hopeless. There is also a lack of a clear decentralized market interpretation in such a non-convex setting.


The approach builds on the work of \cite{davila}, who was able to solve the EVA for all climatic scenarios of the ERAA. This achievement is a considerable stepping stone towards integrating EVA with adequacy assessment, as it allows scaling up the capacity expansion problem for a non-negligible number of scenarios. 
In the present analysis we take a step further towards integration, by adding reliability constraints to the EVA.


\subsection{Paper Structure}

The remainder of the paper is organized as follows: section \ref{Model} describes the model that we aim to solve. Then section \ref{Approach} demonstrates the proposed solution strategy. Section \ref{case_study} presents the results of the proposed approach when applied to a pan-European setting for specific limits on EENS. Finally, section \ref{conclusion} provides concluding remarks. 

\section{Stochastic CEP Model Formulation}
\label{Model}
The CEP model that we will analyze in this paper is an extension of the model used in the work of  \cite{davila} which provided a two-stage stochastic formulation of the EVA model of ERAA 2021 \cite{ERAA21_method}, where the first stage determines investments and retirements, and the second stage optimizes operation for a single year. The extension regards the addition of specific EENS constraints. For the sake of consistency, we have mostly maintained the same nomenclature as \cite{davila}, which is presented at the end of this document. A full list of the equations will be provided here, however, we refer the reader to \cite{davila} for a more detailed description. 


For every zone, the model considers the possibility of retiring existing capacity or investing in new capacity in thermal power generation. Generation from thermal units is limited by available capacity. Must-run obligations of such units are taken into consideration. Maintenance schedules are also accounted for, by using a capacity factor. 


Generation from hydropower units is modelled for four types of hydropower technology per zone: run-of-river, reservoir, open loop pumped storage and closed loop pumped storage.  Battery storage is also modelled at zonal level. It should be noted that capacities in hydropower and battery technologies are not considered as candidates for either investment or retirements; their capacity is exogenously set and the model determines their generation. Finally, the model includes a linearized approximation of the transmission network, with flows being restricted by transfer capacity limits. No investment in transfer capacity is modelled. 


The stochastic parameters of the model are demand, wind and solar PV generation, and hydraulic inflows. For describing the model, we define a scenario as a realization of uncertainty. 


We examine separately the variables and expressions that are scenario-dependent from those that are not. Scenario-independent variables regard the decisions relative to investments and retirements. The cost of those decisions includes the annualized cost of investment, as well as the fixed cost of new operating capacities. It also includes the avoided fixed cost of retired capacities:  
\begin{equation}
\label{eq:first_stage_cost} \sum_{n \in N} \sum_{g \in G^{\nu}} \left ( IC_{n,g} + FOM_{n,g} \right ) \cdot x^{\nu}_{n,g} -\sum_{n \in N} \sum_{g \in G} FOM_{n,g} \cdot x_{n,g} 
\end{equation}
New investments are constrained by an upper limit, while retirements are constrained by existing capacities:
\begin{align}
\label{eq:cap_new} & x^{\nu}_{n,g} \leq X^{\nu}_{n,g}, \;\forall n\in N, g \in G^{\nu}  \\ 
\label{eq:cap_ret} & x_{n,g} \leq X_{n,g}, \;\forall n\in N, g \in G 
\end{align}
The variables that are dependent on the uncertainty realization (or scenario) are those relative to the operation of capacities, transmission flows, storage and load shedding. For every scenario $\omega$ we define the following cost quantities:
\begin{align}
&\sum_{t \in T} \sum_{n \in N} \sum_{g \in G^{\nu}} \Delta T_t \cdot MC_{n,g} \cdot p^{\nu}_{t,n,g,\omega} \nonumber \\
\label{eq:gen_cost}     & + \sum_{t \in T} \sum_{n \in N} \sum_{g \in G} \Delta T_t \cdot MC_{n,g} \cdot p_{t,n,g,\omega}  \\ 
\label{eq:trade_cost}   & \sum_{t \in T} \sum_{n \in N} \sum_{l \in L(n)} \Delta T_t \cdot f_{t,n,l,\omega} \cdot WC \\ 
\label{eq:storage_cost} & \sum_{t \in T} \sum_{n \in N} \sum_{h \in H(n)} \Delta T_t \cdot s_{t,n,h,\omega} \cdot SC  
\end{align}
which represent the cost of power generation (\ref{eq:gen_cost}), the cost of transmission (\ref{eq:trade_cost}) and the cost of water spillage (\ref{eq:storage_cost}). The following constraints apply for every scenario $\omega$:
{\allowdisplaybreaks
\begin{itemize}
\item Power generation constraints for new capacities:
\begin{flalign}
\label{eq:pg_new}  (\mu^{\nu}_{t,n,g,\omega}): &p^{\nu}_{t,n,g,\omega} \leq x^{\nu}_{n,g}, \;\forall t \in T, n\in N, g \in G^{\nu}  &
\end{flalign}
\item Power generation constraints for existing capacities:
\begin{align}
\label{eq:pg_ret1} & (\mu_{t,n,g,\omega}):  &  p_{t,n,g,\omega} \leq P^{max}_{t,n,g} - x_{n,g} &\\
\label{eq:pg_ret2} && p_{t,n,g,\omega} \geq P^{min}_{t,n,g} &
\end{align}
\begin{flushright}
$\forall t \in T, n\in N, g \in G$
\end{flushright}
\item  Transmission network constraints:
\begin{align}
 &f_{t,n,l,\omega} \leq L^{max}_{n,l} &\\
 &f_{t,n,l,\omega} \geq L^{min}_{n,l} & \label{eq: end_of_transmission_constraints}
\end{align}
\begin{flushright}
$\forall t \in T, n\in N, l \in L(n)$
\end{flushright}
\item Battery constraints:
\begin{align}
\begin{split}
bv_{t,n,\omega} = & bv_{t-1,n,\omega} + BCE \cdot bc_{t,n,\omega} \\
& - BDE \cdot bd_{t,n,\omega} \\
\end{split} \\
& bv_{t,n,\omega} \leq BV \\
& bc_{t,n,\omega} \leq BC \\
& bd_{t,n,\omega} \leq BD \label{eq: end_of_battery_constraints}
\end{align}
\begin{flushright}
$\forall t \in T, n\in N, g \in G$
\end{flushright}
\item Run of river constraints:
\begin{equation}
q_{t,n,R,\omega} = A_{t,n,R,\omega}, \; \forall t \in T, n\in N \label{eq: end_of_ror_constraints}
\end{equation}
\item Reservoir constraints:
\begin{align}
\begin{split}
v_{t,n,S,\omega}  = & v_{t-1,n,S,\omega} + A_{t,n,S,\omega} - q_{t,n,S,\omega}  \\
& -s_{t,n,S,\omega} \\
\end{split} \\
& v_{t,n,S,\omega} \leq V_{n,S} \\
& q_{t,n,S,\omega} \leq Q_{n,S} \label{eq: end_of_reservoir_constraints}
\end{align}
\begin{flushright}
$\forall t \in T, n\in N$
\end{flushright}
\item Open loop pumped storage constraints: 
\begin{align}
\begin{split}
 v^{H}_{t,n,O,\omega} = & v^{H}_{t-1,n,O,\omega} + A_{t,n,O,\omega} + PE \cdot d_{t,n,O,\omega} \\
& - q_{t,n,O,\omega} - s_{t,n,O,\omega} \\
\end{split} \\
& v^{H}_{t,n,O,\omega} = v^{H}_{t-1,n,O,\omega} - PE \cdot d_{t,n,O,\omega} + q_{t,n,O,\omega}\\
& v^{H}_{t,n,O,\omega} \leq V_{n,O}  \\
& q_{t,n,O,\omega} \leq Q_{n,O} \\
& s_{t,n,O,\omega} \leq D_{n,O} \label{eq: end_of_open_loop_constraints}
\end{align}
\begin{flushright}
$\forall t \in T, n\in N$
\end{flushright}
\item Closed loop pumped storage constraints:
\begin{align}
\begin{split}
 v^{H}_{t,n,C,\omega} = & v^{H}_{t-1,n,C,\omega} + PE \cdot d_{t,n,C,\omega}  - q_{t,n,C,\omega} \\
 & - s_{t,n,C,\omega} \\
\end{split} \\
 & v^{T}_{t,n,C,\omega} = v^{T}_{t-1,n,C,\omega} - PE \cdot d_{t,n,C,\omega} + q_{t,n,C,\omega} \\
 & v^{H}_{t,n,C,\omega} \leq V_{n,C} \\
 & q_{t,n,C,\omega} \leq Q_{n,C} \\
 & d_{t,n,C,\omega} \leq D_{n,C} \label{eq:end_of_closed_loop_constraints}
\end{align}
\begin{flushright} 
$\forall t \in T, n\in N$
\end{flushright}
\item Load balance
\begin{align}
 \nonumber & \sum_{g\in G}p_{t,n,g,\omega} + \sum_{g\in G^{\nu}}p^{\nu}_{t,n,g,\omega} + bd_{t,n,\omega} + \sum_{h \in H(n)} q_{t,n,h,\omega} \\
 \nonumber & + \sum_{l \in L(n)} f_{t,n,l,\omega} + ls_{t,n,\omega} + PV_{t,n,\omega} + W_{t,n,\omega} \\
 \label{eq:balance} & = D_{t,n,\omega} + bc_{t,n,\omega} + \sum_{r \in \{C,O\}} d_{t,n,r,\omega} 
\end{align}
\begin{flushright} $\forall t \in T, n\in N$ \end{flushright}
\end{itemize}
}

What remain to be defined are the EENS constraints. The EENS constraints bundle all scenarios together, so that for every node of the power system considered, the expected value of load shedding is less than or equal to a predefined EENS limit: 
\begin{align}
\label{eq:EENS} &(\lambda_{n}): & \mathbb{E}_{\omega} \left[  \sum_{t \in T} \Delta T_t \cdot ls_{t,n,\omega} \right] \leq EENS_{n}, \forall n \in N 
\end{align}
Overall, the stochastic capacity expansion model with EENS constraints is formulated as follows:
\begin{align*}
&\min_{x, x^{\nu}, p, p^{\nu}, f, ls} (\ref{eq:first_stage_cost}) + \mathbb{E}_{\omega} \left[ (\ref{eq:gen_cost}) +(\ref{eq:trade_cost}) + (\ref{eq:storage_cost}) \right] \\
\begin{split}
    \text{subject to\;} & (\ref{eq:cap_new}), (\ref{eq:cap_ret}) \\
        & (\ref{eq:pg_new}) - (\ref{eq:balance}), \;\forall \omega \in \Omega \\
        & (\ref{eq:EENS}) \\
\end{split}\\
&\tag{CEP-EENS} \label{CEP-EENS} 
\end{align*}

\section{Proposed Solution}\label{Approach}

\subsection{Lagrange Relaxation Scheme} 
\label{Outer}

In order to solve the \ref{CEP-EENS} problem we will consider relaxing constraints (\ref{eq:EENS}) that are coupling all scenarios. This relaxation will lead us to the following Lagrangian function:
\begin{equation}
\begin{split}
    L(x, & x^{\nu}, p, p^{\nu}, f, s, ls, \lambda)  = (\ref{eq:first_stage_cost}) + \mathbb{E}_{\omega} \left[ (\ref{eq:gen_cost}) +(\ref{eq:trade_cost}) + (\ref{eq:storage_cost}) \right] \\
    & + \sum_{n \in N} \lambda_n \cdot \mathbb{E}_{\omega} \left[  \sum_{t \in T} \Delta T_t \cdot ls_{t,n,\omega}  - EENS_{n} \right] \label{eq:Lagrangian}
\end{split}
\end{equation}
where the Lagrangian multipliers $\lambda = \{ \lambda_n, n \in N\}$ are the shadow prices of constraints (\ref{eq:EENS}) and represent the penalization of the violation of each EENS constraint. We can now define the Lagrange dual function as:
\begin{align*}
g(\lambda) = &\min_{x, x^{\nu}, p, p^{\nu}, f, s, ls} (\ref{eq:Lagrangian}) \\
\begin{split}
    \text{subject to\;} & (\ref{eq:cap_new}), (\ref{eq:cap_ret}) \\
        & (\ref{eq:pg_new}) - (\ref{eq:balance}), \;\forall \omega \in \Omega \\
\end{split}\\
&\tag{Lagrange dual function} \label{Lagrange_dual_function}
\end{align*}

The Lagrange dual function represents the minimum cost capacity expansion plan for a given $\lambda$. From duality theory \cite{boyd2004convex}, $g(\lambda)$ is a lower bound to the \ref{CEP-EENS} problem. By strong duality, the optimal solution to the \ref{CEP-EENS} problem can be attained through: 
\begin{align}
    & \max_{\lambda \geq 0} g(\lambda) \nonumber \tag{Dual CEP-EENS}\label{dual}
\end{align}

To solve the \ref{dual} we propose a subgradient-based algorithmic method \cite{boyd_subgradient} which iterates over trial values of $\lambda$ until a convergence criterion is met. The steps of the process are:
\begin{algorithm}[H]
\caption{Subgradient-based method to solve the Dual CEP-EENS (Main algorithm)}
\begin{algorithmic}
\STATE {\textsc{Input}:} Provide initial values $\lambda^{0} = \{ \lambda^{0}_n, n \in N\}$
\STATE {\textsc{Iteration index}:} $k=0$
\STATE {\textsc{While} Convergence criterion is not met} 
\begin{enumerate}
    \STATE Compute $g(\lambda^{k})$. 
    \STATE Calculate subgradients of $g$ at $\lambda^{k}$, $\rho^{k} = \{ \rho^{k}_n, n \in N\}$
    \STATE Calculate $\lambda^{k+1}$
    \STATE $k \gets k + 1$
\end{enumerate}
\end{algorithmic}
\label{alg1}
\end{algorithm}

Let us now explain the algorithm in more detail. The process begins by providing initial values for $\lambda$, which can take any positive value. Then, the first step of the algorithm is to solve the dual problem for the given $\lambda^{0}$. At every iteration $k$, the value of $g(\lambda^k)$ is a lower bound to our original problem.


In step 2 of the algorithm, we calculate the subgradients of $g$ at $\lambda$. From (\ref{eq:Lagrangian}) we see that these subgradients can be obtained as:
\begin{equation}
    \rho^{k}_n = \mathbb{E}_{\omega} \left[  \sum_{t \in T} \Delta T_t \cdot ls^{k}_{t,n,\omega}\right]  - EENS_{n}, \;\forall n \in N
\end{equation}

Step 3 of the algorithm involves the update of the values of $\lambda$, which, according to the projected subgradient method \cite{boyd_subgradient} are obtained as:
\begin{equation}
    \lambda^{k+1}_n = \max{(0, \lambda^{k}_n + \alpha^{k} \cdot \rho^{k}_n)}, \;\forall n \in N
\end{equation}
The term $\alpha^{k}$ is a step size, for which there are many options to choose from \cite{boyd_subgradient}. In this application we have selected to use the Polyak step size, which ensures convergence \cite{boyd_subgradient}:
\begin{equation}
    \alpha^{k} = \frac{W^{\ast}-W^{k}}{\sum_{n \in N} (\rho^{k}_n)^2}.
\end{equation}
Here, $W^{k}$ is the current solution of our problem (the value of $g(\lambda^{k})$ in our case) and $W^{\ast}$ is the dual optimal value, which can be approximated by an upper bound of the dual optimal value. The estimation of an upper bound is explained in detail in the following subsection \ref{Estimate UB}. 


The algorithm is repeated until the gap between the lower bound (the value of $g(\lambda^{k})$) and the estimated upper bound of the problem is lower than a predefined target. 


Steps 2 and 3 of the proposed algorithmic process are trivial from a computational standpoint. What is non-trivial is the first step, as it amounts to solving a large stochastic capacity expansion problem. To tackle this, we implement the methodology introduced in \cite{davila}, which is described in section \ref{Inner}.

\subsection{Estimating an Upper Bound} 
\label{Estimate UB}

For every $\lambda$ (except for the one that maximizes $g(\lambda)$), the solution of $g(\lambda)$ satisfies all constraints of the original \ref{CEP-EENS} problem, except for the EENS constraints. In other words, a solution to $g(\lambda)$ is typically infeasible to the original problem, as there is excess load shedding in some nodes. 


Assume that we are able to add capacity to the solution of $g(\lambda)$ which would be sufficient to cover for the excess load shedding. Such an expansion decision 
would yield a feasible solution to the \ref{CEP-EENS} problem, which would allow us to compute an upper bound to the problem,  since the \ref{CEP-EENS} is a minimization problem. With this in mind, we have developed a ``feasibility recovery" procedure, which, given a solution of $g(\lambda)$, computes the required additional investments in a specific type of capacity (specifically peaking units), so that the EENS constraints (\ref{eq:EENS}) are respected at every node. 


The procedure is depicted in pseudocode in Algorithm \ref{algfeas}. It considers separately every node for which the EENS constraint is violated. The limit violation in every one of these “infeasible” nodes is interpreted as a generation deficit. At every iteration of the algorithm, the deficit of generation is gradually reduced by introducing additional capacity, until the deficit is entirely eliminated.


In more detail, the algorithm begins by considering an initial deficit of generation (equal to the limit violation, denoted as $Deficit$) as well as the expected number hours for which load shedding occurs (denoted as $LOLE$). It then calculates the amount $x^+ = \frac{Deficit}{LOLE}$ which is the least amount of capacity that is required to cover for the deficit of generation. Note that if load shedding were equally spread across hours, this amount of capacity would be sufficient to cover all of the deficit.


The algorithm proceeds with calculating reduced values for load shedding $ls^{f}$, assuming that the $x^+$ that has been added can generate up to $x^+$ (accounting also for a capacity factor, if applicable) during load shedding hours. Considering the new values of load shedding, the algorithm recalculates the deficit, which will be reduced. The process is repeated until the deficit reaches zero, and thus $ls^{f}$ is a feasible amount of load shedding.

\begin{algorithm}[H]
\caption{Find missing capacity for feasibility recovery}
\begin{algorithmic}
\STATE {\textsc{Input}:} Load shedding values $\hat{ls}$ from solution of $g(\lambda^{k})$ for a particular node, with limit violation of node 
$\mbox{LimitViolation} = \mathbb{E}_{\omega} \left[  \sum_{t \in T} \Delta T_t \cdot \hat{ls}_{t,n,\omega} \right] - EENS_n \geq 0$.
\STATE {\textsc{Initializations:}} $ls^{f} = \hat{ls}$, $\mbox{Deficit} = \mbox{LimitViolation}$, $x^{new}=0$
\STATE {\textsc{While} true} 
\begin{enumerate}
\STATE $x^+ = \frac{\mbox{Deficit}}{\mbox{LOLE}}$
\STATE  $x^{new} \gets x^{new} + x^+$
\STATE $ls^{f} \gets \max\{ls^{f} - x^+, 0\}$
\STATE $\mbox{Deficit} \gets \mathbb{E}_{\omega} \left[  \sum_{t \in T} \Delta T_t \cdot ls^{f}_{t,n,\omega} \right] - EENS_n$
\IF{$\mbox{Deficit}\leq 0$} 
    \STATE  End. Have found $x^{new}$
\ENDIF
\end{enumerate}
\end{algorithmic}
\label{algfeas}
\end{algorithm}

An approach towards the same objective could be to fix the level of investments as found by minimizing $g(\lambda^{k})$, except for specific technologies, and solve the \ref{CEP-EENS}. However, even with some fixed investments, solving the \ref{CEP-EENS} would be hard for a large number of scenarios and nodes. Algorithm \ref{algfeas} is a heuristic alternative, the advantage of which lies exactly in that we avoid solving a large-scale problem that would delay our iterative process to an unacceptable level.


Having such a procedure that augments capacity in each node so that the EENS constraints are satisfied allows us to calculate an upper bound to our problem. The following proposition formulates this argument mathematically.


\textbf{Proposition}: Consider $\lambda\geq 0$ and $\{\hat{x}, \hat{x}^{\nu}, \hat{p}, \hat{p}^{\nu},\hat{f}, \hat{s},  \hat{ls}\}$ the solution obtained when computing $g(\lambda)$. There exist $x^f, p^f$ such that the following inequality holds:
\begin{equation*}
\ref{CEP-EENS} \leq UB(\lambda) 
\end{equation*}
where
\allowdisplaybreaks{
\begin{equation*}
\begin{split}
    UB(\lambda) = & \sum_{n \in N} \sum_{g \in G^{\nu}} \left ( IC_{n,g} \cdot x^f_{n,g} + FOM_{n,g} \cdot x^f_{n,g} \right) \\
     & -\sum_{n \in N} \sum_{g \in G} FOM_{n,g} \cdot \hat{x}_{n,g}  \\ 
     &  + \mathbb{E}_{\omega} \left[ \sum_{t \in T} \sum_{n \in N} \sum_{g \in G^{\nu}} \Delta T_t \cdot MC_{n,g} \cdot p^f_{t,n,g,\omega} \right]  \\ 
     & + \mathbb{E}_{\omega} \left[ \sum_{t \in T} \sum_{n \in N} \sum_{g \in G} \Delta T_t \cdot MC_{n,g} \cdot \hat{p}_{t,n,g,\omega} \right]   \\ 
     & + \mathbb{E}_{\omega} \left[ \sum_{t \in T} \sum_{n \in N} \sum_{l \in L(n)} \Delta T_t \cdot \hat{f}_{t,n,l,\omega} \cdot WC \right ]  \\ 
     & + \mathbb{E}_{\omega} \left[ \sum_{t \in T} \sum_{n \in N} \sum_{h \in H(n)} \Delta T_t \cdot \hat{s}_{t,n,h,\omega} \cdot SC \right ]      
\end{split}  
\end{equation*}}

\textbf{Proof}: Without loss of generality, we will consider that all nodes are infeasible in the solution of $g(\lambda)$ and that the upper limit on investments ($X_{n,g}$) is high enough so that feasibility recovery will not violate capacity constraint (\ref{eq:cap_new}) (step 2 of Algorithm \ref{algfeas}). 
Let $\mathcal{C}$ be the set of hours of curtailment incidents. Let $\tilde{g}\in G$ be the type of generator whose capacity will be augmented to recover feasibility.
Let $x^{new}$ and $ls^f$ be the feasibility recovery capacity and the feasible load shedding respectively, as provided by Algorithm \ref{algfeas} for every $n \in N$. Then, we define a new set of investments: 
\begin{equation}
  x_{n,g}^{f} =
    \begin{cases}
      \hat{x}^{\nu}_{n,g} & \text{ if } g\neq\tilde{g}\\
      \hat{x}^{\nu}_{n,g} + x^{new}_n & \text{ if } g = \tilde{g}\\
    \end{cases}  \nonumber     
\end{equation}
We also define a new generation profile: 
\begin{equation}
  p_{t,g,\omega}^{f} =
    \begin{cases}
      \hat{p}^{\nu}_{t,g,\omega} & \text{ if } g\neq\tilde{g}\\
      \hat{p}^{\nu}_{t,g,\omega} & \text{ if } g = \tilde{g} \text{ and } t\notin\mathcal{C}\\
      \hat{p}^{\nu}_{t,g,\omega} + (\hat{ls}_{t,\omega} -ls_{t,\omega}^{f}) & \text{ otherwise}
    \end{cases}  \nonumber     
\end{equation}
The proposition is true if $\{x^{f}, \hat{x}, p^{f}, \hat{p}, \hat{f}, \hat{s}, ls^{f}\}$ is a feasible solution to the \ref{CEP-EENS} problem. Indeed, constraints (\ref{eq:cap_ret}), (\ref{eq:pg_ret1}) - (\ref{eq:end_of_closed_loop_constraints}) are satisfied because $\{\hat{x}, \hat{p},\hat{f}, \hat{s}\}$, are part of the solution to $g(\lambda)$ (which satisfies all of the \ref{CEP-EENS} problem constraints, except for the EENS constraints). Moreover, the termination criterion of Algorithm \ref{algfeas} ensures that $ls^f$ are such that the EENS constraints (\ref{eq:EENS}) are satisfied. Furthermore, the definition of $p^f$ is such that constraints  (\ref{eq:pg_new}) and (\ref{eq:balance}) are also satisfied. Finally, it has been assumed that constraints (\ref{eq:cap_new}) are also satisfied. 
Therefore, it follows that $\ref{CEP-EENS} \leq UB(\lambda)$. $\blacksquare$


Note that the above proposition holds for any type of capacity $g \in G^{\nu}$. With small modifications in step 2 of Algorithm \ref{algfeas}, it can be ensured that additional capacity does not exceed the upper limits of constraint (\ref{eq:cap_new}). We can thus select to cover the excess load shedding with more than one type of capacity. In this case, Algorithm \ref{algfeas} would be implemented sequentially for each. 
Retirements can also be considered along with new investments. In our implementation, we have chosen to first recover retired capacity at every node and then, if load shedding remains above limits after recovering all retirements, to add new investments in peaking capacity. A similar approach was adopted in the ERAA 2021 study \cite{ERAA21_method} for the "Scenario with CM".

\subsection{Decomposition Scheme for Computing the Lagrange Dual Function} \label{Inner}

As described in the previous section, our approach requires that, at every iteration $k$, we compute the Lagrange dual function for a specific set of Lagrange multipliers $\lambda^k, g(\lambda^k)$. 
This is a capacity expansion problem for a specific  penalization of load shedding $\lambda^k$. In order to solve it, we rely on the work of \cite{davila}, who propose to implement a subgradient based scheme also in this step.

\par
Let us slightly reformulate the problem that we need to minimize. We will name it \ref{CEP-EENS-R}, to denote that it comes from the Lagrange relaxation of the EENS constraints of the original \ref{CEP-EENS} problem. The constant terms $\lambda_n\cdot EENS_n$ can be removed from $L$ (\ref{eq:Lagrangian}) since they do not affect the minimization. 
We are thus required to solve the problem:
\begin{align*}
&\min_{x, x^{\nu}, p, p^{\nu}, f, s, ls} \\
&(\ref{eq:first_stage_cost}) + \mathbb{E}_{\omega} \left[ (\ref{eq:gen_cost}) +(\ref{eq:trade_cost}) + (\ref{eq:storage_cost}) + \sum_{n \in N} \lambda_n \sum_{t \in T} \Delta T_t \cdot ls_{t,n,\omega} \right] \\
\begin{split}
    \text{subject to\;} & (\ref{eq:cap_new}), (\ref{eq:cap_ret}) \\
        & (\ref{eq:pg_new}) - (\ref{eq:balance}), \;\forall \omega \in \Omega \\
\end{split} \\
&\tag{CEP-EENS-R} \label{CEP-EENS-R}
\end{align*}
Problem \ref{CEP-EENS-R} has a distinct structure of two-stage optimization under uncertainty.  
The first stage regards the investment variables $\{x^{\nu}, x\}$, which are decided prior to the realization of uncertainty; the second stage regards all other operation-related variables that assume a different value for every uncertainty realization. Second-stage variables depend on the value of first-stage variables. This formulation allows us to decompose the problem. In particular, for a given level $\{x^{\nu}, x\}$ we may define a second-stage subproblem (which allows us to compute a value function) for each scenario $\omega$:

\begin{align*}
V(x^{\nu},x,\omega) = & \min_{p, p^{\nu}, f, s, ls} (\ref{eq:gen_cost}) +(\ref{eq:trade_cost}) + (\ref{eq:storage_cost}) \\
    & + \sum_{n \in N} \lambda_n \sum_{t \in T} \Delta T_t \cdot ls_{t,n,\omega} \\
    & \text{subject to\;} (\ref{eq:pg_new}) - (\ref{eq:balance}) \\
&\tag{Value functions} \label{eq:vxw}
\end{align*}

These subproblems are relatively easy to solve since $\{x^{\nu}, x\}$ are fixed and each subproblem corresponds to a single scenario. We can finally rewrite the problem in its new form, named CEP-EENS-RD (adding D for ``decomposition"):
\begin{align*}
    &\min_{x^{\nu},x} (\ref{eq:first_stage_cost}) + \mathbb{E}_{\omega} \left[ V(x^{\nu},x,\omega) \right] \\
    & \text{subject to\;} (\ref{eq:cap_new}), (\ref{eq:cap_ret}) \\
    &\tag{CEP-EENS-RD} \label{CEP-EENS-RD} 
\end{align*}

The approach suggested by \cite{davila} to tackle this problem is to find subgradients of the objective function along the coordinates of $x^{\nu}$ and $x$, and make updates along such a direction. Consider a candidate expansion plan $\{{x^{\nu}}^i, x^i\}$ and a solution to the second-stage subproblems $V({x^{\nu}}^i,x^i,\omega)$, from which we obtain the dual multipliers of the constraints (\ref{eq:pg_new}) and (\ref{eq:pg_ret1}), $\mu^{\nu}$ and $\mu$ respectively. Then, the subgradients along $\{x^{\nu}, x\}$ can be obtained from: 
\begin{align}
\label{eq:q1} & q^{\nu}_{n,g} = IC_{n,g} + FOM_{n,g} + \mathbb{E}_{\omega}\left[ \sum_{t \in T}\mu^{\nu}_{t,n,g,\omega} \right] \\
\label{eq:q2} & q_{n,g} = - FOM_{n,g} + \mathbb{E}_{\omega}\left[ \sum_{t \in T}\mu_{t,n,g,\omega} \right]
\end{align}
Using a projected subgradient method, the candidate expansion plan can thus be updated as: 
\begin{align}
\label{eq:x1} & {x^{\nu}}^{i+1} = \min\{X^{\nu}_{n,g}, {x^{\nu}}^i_{n,g}-\alpha^i \cdot {q^{\nu}}^i_{n,g}\}  \\
\label{eq:x2} & x^{i+1} = \min\{X_{n,g}, x^i_{n,g}-\alpha^i \cdot q^i_{n,g}\}
\end{align}
Similarly to the subgradient scheme that was presented in section \ref{Outer}, the Polyak stepsize is chosen as:
\begin{equation}
 \label{eq:a2}   \alpha^{i} = \frac{W^{i}-W^{\ast}}{\sum_{n \in N, g \in G^{\nu}} ({q^{\nu}}^{i}_{n,g})^2 + \sum_{n \in N, g \in G} (q^{i}_{n,g})^2} 
\end{equation}
where $W^i$ is the current iterate objective value of \ref{CEP-EENS-RD}, and $W^{\ast}$ is the optimal value.

\par

Since $W^{\ast}$ is not known in advance, an estimate can be used \cite{boyd_subgradient}. Such an estimate is the ``wait-and-see" solution of the \ref{CEP-EENS-R}, which is a lower bound to the true $W^{\ast}$ \cite{birge2011introduction}. To obtain the wait-and-see solution, we solve \ref{CEP-EENS-R}  separately for every scenario. If we define as $f^{WS}_{\omega}$ the objective value of these scenario-specific problems, then we obtain $W^{\ast}$ as $\mathbb{E}_{\omega}\left[ f^{WS}_{\omega} \right]$.

\par

The steps for solving the problem are summarized in the following algorithm, to which we will be referring to as the ``inner algorithm": 
\begin{algorithm}[H]

\caption{Subgradient-based method to solve the \ref{CEP-EENS-RD} (Inner algorithm)}
\begin{algorithmic}
\STATE {\textsc{Input}:} Provide initial values ${x^{\nu}}^0, x^0$ and an estimate $W^{\ast}$
\STATE {\textsc{Iteration index}:} $i=0$
\STATE {\textsc{While} Convergence criterion is not met} 
\begin{enumerate}
    \STATE Solve $V({x^{\nu}}^i,x^i,\omega)$ for every $\omega \in \Omega$ and retrieve  $\mu^{\nu}$,$\mu$ 
    \STATE Calculate subgradients $q^{\nu}_{n,g}$, $q_{n,g}$ according to Eqs. (\ref{eq:q1}), (\ref{eq:q2})
    \STATE Update ${x^{\nu}}^{i+1}, x^{i+1}$ according to Eqs. (\ref{eq:x1}) - (\ref{eq:a2})
    \STATE $i \gets i + 1$
\end{enumerate}
\end{algorithmic}
\label{alg3}
\end{algorithm}
 
The convergence criterion of the inner algorithm has been selected to be the stabilization of the objective of the current iterate.

\subsection{Parallelization Strategy} 
The proposed scheme includes two processes that require the solution of scenario-specific subproblems. The first one is obtaining a lower bound for the Polyak stepsize for the inner algorithm (\ref{eq:a2}). The second one is step 1 of the inner algorithm (Algorithm \ref{alg3}). The scenario-specific problems are independent to each other, they can thus be solved in parallel. We can therefore implement the proposed scheme in a distributed computation infrastructure that can handle the parallelization.

\section{Case Study}\label{case_study}

\subsection{Case Study Description}

In this section we demonstrate results of the methodology for a system that is based on the ERAA 2021 study \cite{ERAA21_method}, which analyzes 56 nodes (each corresponding to a bidding zone) of Europe, for 35 climatic scenarios. Necessary input data to populate the model \ref{CEP-EENS} and to create the 35 scenarios are available from the ERAA website \cite{eraa_downloads}. In cases that necessary input information was not available (e.g. marginal costs of units), typical values are used.


\subsubsection{EENS Limits} 

In what follows we present results for different cases of EENS limits by country, however the basic case that we are analysing uses the EENS limits from the respective results of ERAA 2021 for year 2025, for the scenario "EVA with CM" (Table 8 of \cite{ERAA21_results}). We refer to this basic case as "ERAA21". Other EENS cases that we investigate correspond to fractions of annual demand. In case ``Low", limits are set to 0.01$\%$ of the minimum annual demand across scenarios. The same figure is 0.1$\%$ in case ``High". Finally, in case ``Zero", EENS limits are set to zero.


\subsubsection{Benchmark approach}

For benchmarking the results of the algorithm, we compare them to the results we would obtain if instead of reliability constraints we used a load shedding penalization (VOLL or price cap), as implemented in the EVA. In other words, we compare our approach that explicitly considers reliability constraints with an approach where such constraints are considered implicitly. For this purpose, we perform a single run of the algorithm setting $\lambda$ at a load shedding penalization value that is expected to yield reliability indices similar to the imposed limits. 
The basic case "ERAA21" is used for this benchmarking, as a consistent value of load shedding penalization is known to us; this is the load shedding penalization that was used in the EVA of ERAA 2021, at 15 000 EUR/MWh \cite{ERAA21_assumptions}.  

\subsubsection{Algorithm configuration}
We set the value of initial $\lambda$ to be very low (50 EUR/MWh) for all bidding zones. We are thus quite far from what we expect the optimal solution to be. This allows us to observe the full evolution of the algorithm. We allow the algorithm to evolve until a satisfactory optimality gap is achieved. The feasible investment plan that corresponds to the lowest upper bound (see \ref{Estimate UB}) is retrieved as the final investment plan. Ex-post, we fix the level of investments to this feasible plan and solve the \ref{CEP-EENS-R} problem to obtain its true cost.

 
For the inner algorithm (Algorithm \ref{alg3}), we set the termination criterion to a 0.1$\%$ relative change of the objective value over consecutive iterations.  
We acknowledge that this criterion offers no convergence guarantees, but through experiments on smaller-scale systems for which we could compute the extended form solution, it was empirically observed to exhibit satisfactory behavior.


\subsubsection{Computational environment}

We use the Julia programming language (v1.6.5) to implement the algorithms. Paralellization of the work has been achieved through packages Parallelism (v0.1.3) and Distributed. Relevant packages used for building the models are JuMP (v0.21.10) and ParameterJuMP (v0.3.0). Models are solved using the Gurobi 9 solver. All runs are implemented on GRNET’s HPC resources ``ARIS", on nodes with 4 SandyBridge - Intel(R) Xeon(R) CPU E5-4650v2 processors, with 2.4 GHz frequency and 10 cores each. 


Running the application requires a minimum of 35 CPUs, one for each scenario/parallel task. The process can be run with a higher number of CPUs per parallel task, through employing parallel algorithms of commercial solvers such as Gurobi (e.g. Barrier) that can mobilize multiple threads for the same process, and accelerate it. The results presented for this case study have been obtained with the most conservative configuration of 1 CPU per task.

\subsection{Performance of the Algorithm}

\begin{figure}[!t]
\centering
\includegraphics[width=12cm]{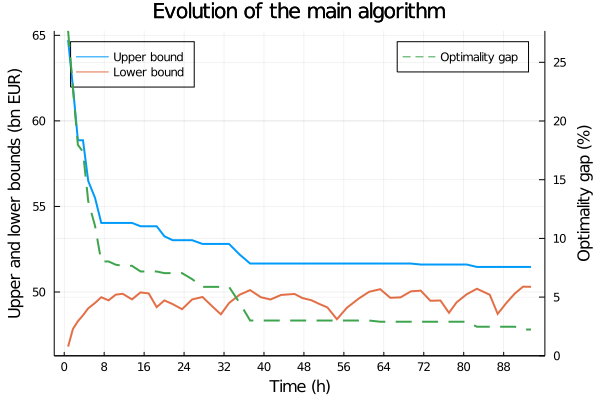}
\caption{Evolution of the main algorithm for the ERAA21 case, starting from an unfavourable initial point and utilizing 1 CPU per task. The optimality gap shown in the dashed line is the best gap obtained so far.}
\label{convergence}
\end{figure}

Fig. \ref{convergence} demonstrates the evolution of the upper and the lower bounds of the main algorithm of the process (Algorithm \ref{alg1}), as well as the optimality gap, for the basic case (ERAA21). With the conservative configuration of 1 CPU per task, a total of 53 iterations were completed in a period of four days. On average, every iteration of the main algorithm required 100 minutes; initialization of the inner algorithm required, on average, 38 minutes. The inner algorithm required an average of 5 iterations, with each iteration taking up approximately 15 minutes. A higher number of CPUs per task would result in faster evolution. For instance, it has been observed that using 5 CPUs per task reduces the duration of each inner algorithm iteration by 70$\%$.


The best optimality gap obtained was 2.2$\%$ in the 52\textsuperscript{nd} iteration. Evaluated ex-post, the true gap (calculated considering the true cost of the final investment plan) is 1.3$\%$.


\subsection{Results for the ERAA21 Case} 

\begin{figure}[!t]
\centering
\includegraphics[width=12cm]{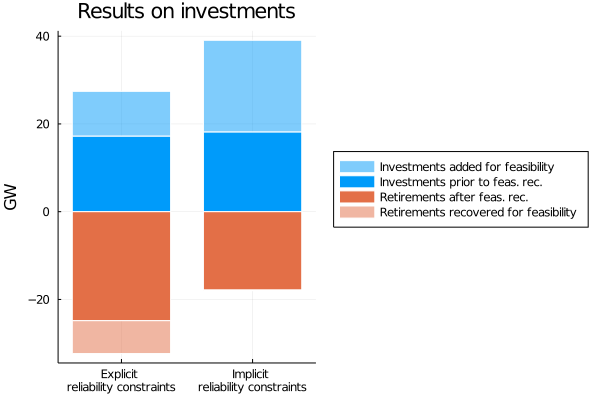}
\caption{Investments results for the case with ERAA21 limits. The case of explicit reliability constraints corresponds to the results of the proposed algorithm, while the implicit case corresponds to results obtained from a single run of the algorithm for a uniform $\lambda$ of 15 000 EUR/MWh for all bidding zones. Final investments and retirements are the ones obtained from the feasibility recovery (feas. rec.) procedure of section \ref{Estimate UB}.
}
\label{fig_Inv_1}
\end{figure}

The final investment plan includes approximately 24 GW of retirements and 27 GW of new investments (aggregate figures for all nodes). These figures are presented in Fig. \ref{fig_Inv_1} under "Explicit reliability constraints", along with the respective figures obtained from the benchmark approach, under "Implicit reliability constraints".


It is apparent from Fig. \ref{fig_Inv_1} that the implicit approach increases investments by approximately 40$\%$. 
Moreover, the implicit approach leads to a solution that requires greater capacity adjustments to ensure feasibility. It appears that the flexibility to adjust the $\lambda$ by bidding zone, which is possible in the proposed approach, allows for meeting the same EENS targets with fewer investments and thus improved cost efficiency. In particular, the total investment cost decreases by 38$\%$; total cost decreases by 1.6$\%$. 

\begin{figure}[!t]
\centering
\includegraphics[width=12cm]{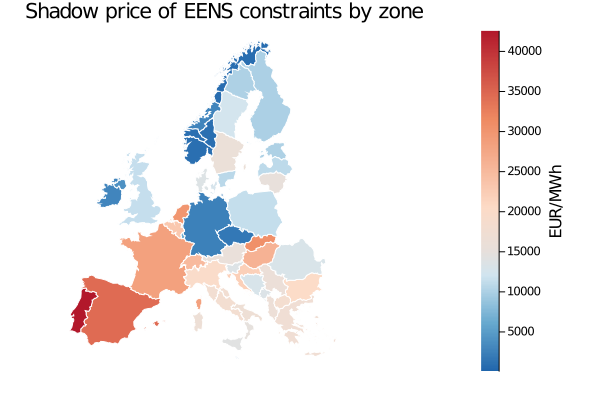}
\caption{The shadow price of EENS constraints ($\lambda$) for the ERAA21 case. Blue is used for values below the benchmark of 15 000 EUR/MWh, and red for values above the benchmark.}
\label{fig_lambda}
\end{figure}

The variation in country-specific $\lambda$ is actually quite significant in the ERAA21 case, as shown in Fig. \ref{fig_lambda}.
Note that the average value of $\lambda$ for all bidding zones is 15 482 EUR/MWh (weighted by average annual demand), which is notably close to the benchmark $\lambda$ of the implicit case. This may serve as an indication of consistency between the level of load shedding penalization and the reliability targets used in ERAA 2021. 

\subsection{Results for Various Cases of EENS Limits}

The proposed algorithm allows us to compare different cases of EENS limits, which in turn can be useful for analysing relevant policies. Let us consider the case were zero load shedding is allowed. The amount of investments required rises to an aggregate of 58 GW, and retirements are limited to 17 GW. By comparing to the ERAA21 case, we see that allowing load shedding that is consistent to the EENS limits of the ERAA 2021 reduces investment requirements by 30 GW at a European level (or 53$\%$), while it allows for retiring an additional 7.3 GW of existing capacity (or 42$\%$). In terms of costs, these figures translate to a 62.7$\%$ reduction in investment cost and a 5$\%$ reduction  in total costs.


Table \ref{Case_results} summarizes the results that are obtained for all four cases that are considered. We observe that there is a significant cost benefit 
from allowing even small amounts of load shedding, as in the ``Low" case. Beyond that point, cost reduction is almost linear with higher amounts of load shedding, although additional cases should be analyzed to concretely comment on the curvature of cost relative to EENS limits.    

\begin{table}[!t]
\caption{Results for different cases of EENS limits.}
\label{Case_results}
\centering
\renewcommand{\arraystretch}{1.3}
\begin{minipage}{\textwidth}
\begin{tabular}{|c|c|c|c|c|}
\hline
EENS limits case & Zero & Low & ERAA21 & High\\
\hline
Aggregate EENS (MWh) & 0 & 17 000 & 125 510 & 191 600  \\
\hline
Average $\lambda$\footnote{weighted by average annual demand} (EUR/MWh) &385 321& 71 605& 14 127& 7738 \\
\hline
Investments (GW)  & 57.9& 40.5& 27.4& 21.4 \\
\hline
Retirements (GW) & 17.5& 18.9& 24.8& 25.5 \\
\hline
Investment cost & & & & \\ 
relative to case Zero ($\%$) & - & -35.2$\%$& -62.7$\%$& -73.9$\%$ \\
\hline
Total cost\footnote{power generation cost plus cost of load shedding valued at $\lambda$} & & & &\\ 
relative to case Zero & - & -1.6$\%$& -5.0$\%$& -6.3$\%$ \\
\hline
\end{tabular}
\end{minipage}
\end{table}

\section{Conclusion}
\label{conclusion}
The current paper presents a methodology for solving a large, stochastic capacity expansion problem with explicit EENS constraints. Given the computational complexity of the task, a common approach in similar studies, in both the literature and real-world cases, is, instead of setting specific reliability limits, to include a penalization of load shedding in the objective function. This penalization is an estimation of the ``true" VOLL. Our analysis demonstrates that this leads to a capacity expansion plan with higher costs relative to the plan that is obtained through our proposed approach. In particular, for the case study of the European power system with EENS targets equal to the ERAA 2021 results, the proposed approach achieves 1.6$\%$ savings in total cost. This is because the proposed approach penalizes load shedding at the value of the dual multipliers of the EENS constraints, which accurately captures the respective cost in every bidding zone.


The proposed approach allows us to observe how imposing stricter or looser EENS limits impacts investments and costs. It also allows us to quantify the benefit of allowing some load shedding in the capacity planning process, e.g. in the case study analyzed allowing for load shedding consistent to the ERAA 2021 EENS results saves $5\%$ in total costs.  


In general, the fact that the proposed approach allows us to retrieve the dual multipliers (or shadow prices) of the EENS constraints is a considerable benefit. Such information can be useful from a policy perspective, as it provides an indication of load shedding penalization that would signal the appropriate level of investments in accordance with a specific EENS limit. Moreover, this information is obtained by bidding zone, while considering the interaction of bidding zones with the interconnected power system. 


In closing, let us mention some points that should be further explored in future work. Initializing the inner algorithm is rather time-consuming, taking up 40$\%$ of the total time. It would thus be beneficial to work on accelerating or eliminating this step.
Moreover, the possibility to expand the approach to a multi-year optimization can be explored. Finally, there is room for studying the policy-relevant aspects that arise from the possibility to obtain zone-specific load shedding penalization values.

\section*{Nomenclature}

\begin{itemize}[wide=0pt]
\item \textbf{Sets and respective indices} 
 \begin{itemize}[wide=0pt]
 \item[] $\Omega, \omega$ : uncertainty realizations
 \item[] $N, n$: bidding zones 
 \item[] $G, g$: existing generators
 \item[] $G^\nu, g$: new generators
 \item[] $T, t$: time horizon
 \item[] $L(n), l$: lines in zone $n$
 \item[] $H(n), h$: hydro technologies in zone $n$, including run-of-river ($R$), reservoir ($S$), open loop ($O$) and closed loop ($C$) pumped storage
\end{itemize}
\item \textbf{Parameters}
    \begin{itemize}[wide=0pt]
    \item[] $IC_{n,g}$: annualized investment cost (€/MW)
    \item[] $FOM_{n,g}$: fixed operational and maintenance cost (€/MW)
    \item[] $MC_{n,g}$: variable and fuel cost (€/MW)
    \item[] $WC$: wheeling charge cost (€/MW)
    \item[] $SC$: water spillage  cost (€/MW)
    \item[] $\Delta T_t$: duration of time block (hours)
     \item[] $X_{n,g},X^\nu_{n,g}$: upper limit for retirement/investment (MW)  
     \item[] $P_{t,n,g}^{max},P_{t,n,g}^{min}$: limits on generation (MW)
     \item[] $L_{n,l}^{max},L_{n,l}^{min}$: limits on transmission flow (MW)
     \item[] $BCE, BDE$: battery charge/discharge efficiency ($\%$)
     \item[] $BV$: battery capacity (MW)
     \item[] $BC$: maximum charge capacity (MW)
     \item[] $BD$: minimum discharge capacity (MW)
     \item[] $V_{n,h}$: maximum storage of hydro unit (MWh)
     \item[] $Q_{n,h}$: turbine capacity of hydro unit (MW)
     \item[] $D_{n,h}$: pump capacity of pumped storage unit (MW)
     \item[] $PE$: pump efficiency ($\%$) 
     \item[] $EENS_n$: EENS limit (MWh)
    \end{itemize}
\item \textbf{Parameters with uncertainty} 
    \begin{itemize}[wide=0pt] 
    \item[] $A_{t,n,h,\omega}$: hydro inflow (MW) 
        \item[] $PV_{n,t,\omega}$: PV production (MW) 
        \item[] $W_{n,t,\omega}$: wind production (MW)
        \item[] $D_{n,t,\omega}$: demand (MW)
    \end{itemize}
\item \textbf{Variables} 
    \begin{itemize}[wide=0pt]
    \item[] $x_{n,g}^\nu$: invested capacity (MW)
    \item[] $x_{n,g}$: retired capacity (MW)
    \item[] $p_{t,n,g,\omega}$: generation by existing capacities (MW)
    \item[] $p_{t,n,g,\omega}^\nu$: generation by new capacities (MW)
    \item[] $f_{t,n,l,\omega}$: power flow (MW)
    \item[] $s_{t,n,h,\omega}$: the amount of water spillage (MW)
    \item[] $ls_{t,n,\omega}$: load shedding (MW)
    \item[] $bv_{t,n,\omega}$: battery state of charge (MWh)
    \item[] $bc_{t,n,\omega}$: power charging the battery (MW)
    \item[] $bd_{t,n,\omega}$: power discharging from the battery (MW)
    \item[] $q_{t,n,h,\omega}:$ the hydro-turbined produced power (MW)
    \item[] $v_{t,n,h,\omega}:$ state of storage of reservoir (MWh)
    \item[] $v_{t,n,h,\omega}^H:$ state of storage of head reservoir (MWh)
    \item[] $v_{t,n,h,\omega}^T:$ state of storage of tail reservoir (MWh)
    \item[] $d_{t,n,h,\omega}:$ pumped power (MW) 
    \end{itemize}

\end{itemize}

\section*{Acknowledgment}
The work has received funding from the European Research Council (ERC) under the European Unions Horizon 2020 research and innovation programme (grant agreement No. 850540).
Moreover, the work was supported by computational time granted from the National Infrastructures for Research and Technology S.A. (GRNET S.A.) in the National HPC facility - ARIS - under project ID pr015023$\_$thin-ICEBERG.

\bibliographystyle{IEEEtran}
\bibliography{CapEx_EENS}

\end{document}